\documentstyle[12pt,aaspp4,flushrt]{article}

\def\chlap#1{\hbox to 50pt{\hss#1\hss}} 
\def\chhlap#1{\hbox to 24pt{\hss#1\hss}} 
\def\chhhlap#1{\hbox to 15pt{\hss#1\hss}} 
\def\alii{{\rm Al}\thinspace{\sc{ii}}}
\def\ang{\thinspace{\rm \AA}}
\def\approxlt{\lower.2em\hbox{$\buildrel < \over \sim$}}
\def\approxgt{\lower.2em\hbox{$\buildrel > \over \sim$}}
\def\cii{{\rm C}\thinspace{\sc{ii}}}
\def\civ{{\rm C}\thinspace{\sc{iv}}}
\def\cmsq{~{\rm cm^{-2}}}
\def\etal{{\it et~al.\/}}
\def\hi{{\rm H}\thinspace{\sc{i}}}
\def\hnought{\ifmmode H_0
    \else $H_0$\fi}
\def\kms{~{\rm km\ s}^{-1}}
\def\la{\ifmmode {{\rm Ly}\alpha}
        \else {Ly$\alpha$}\fi}
\def\lb{\ifmmode {{\rm Ly}\beta}
        \else {Ly$\beta$}\fi}
\def\mgii{{\rm Mg}\thinspace{\sc{ii}}}
\def\nh{\ifmmode {{N_{\rm H}}}
    \else {${N_{\rm H}}$}\fi}
\def\nhi{\ifmmode {{N_{\rm H\thinspace{I}}}}
    \else {${N_{\rm H\thinspace{I}}}$}\fi}
\def\qnought{\ifmmode q_0
    \else $q_0$\fi}
\def\silii{{\rm Si}\thinspace{\sc{ii}}}
\def\siliii{{\rm Si}\thinspace{\sc{iii}}}
\def\siliv{{\rm Si}\thinspace{\sc{iv}}}
\def\ten#1{\ifmmode 10^{#1}
    \else $10^{#1}$\fi}
\slugcomment{\it To appear in the October 1995 Astronomical Journal}

\begin{document}

\title{The Distribution of Column Densities and {$\boldmath b$} Values in the
  Lyman-Alpha Forest}
\author{Esther M. Hu\altaffilmark{1}, Tae-Sun Kim\altaffilmark{1},
  Lennox L. Cowie\altaffilmark{1}, and Antoinette Songaila\altaffilmark{1}}
\affil{Institute for Astronomy, University of Hawaii, 2680 Woodlawn Dr.,
  Honolulu, HI 96822\\
  hu@ifa.hawaii.edu, tsk@ifa.hawaii.edu, cowie@ifa.hawaii.edu,
  acowie@ifa.hawaii.edu}
\and
\author{Michael Rauch}
\affil{Observatories of the Carnegie Institution of Washington, 813 Santa
  Barbara St., Pasadena, CA 91101\\
  mr@ociw.edu}
\altaffiltext{1}{Visiting Astronomer, W. M. Keck Observatory, jointly
  operated by the California Institute of Technology and the University of
  California.}

\begin{abstract}
We describe the properties of the \la\ forest in the column density range
$\nhi \geq 2\times\ten{12}\cmsq$ based on 1056 lines in the wavelength
range $4300-5100\ang$ measured in extremely high S/N, $R=36,000$ spectra
of four quasars.  The column density distribution is well described by a
$-1.5$ power law to $2\times\ten{12}\cmsq$, below which limit confusion
becomes too severe to measure a spectrum of individual clouds.  The
distribution of $b$ values shows a well-defined lower envelope with a
cutoff at $b=20\kms$ corresponding to a cloud temperature of 24,000~K.
There is only a very small fraction (less than 1\%) of narrow line clouds
which cannot be identified with metal-lines.  From modeling the \la\
absorption lines as complexes of clouds each with thermal broadening
corresponding to $b_c$ we find the $b$ distribution can be understood
if there is a mean of 3.25 clouds per absorption line with a spread in
velocity centroids characterized by a dispersion of $10.75\kms$.

\end{abstract}


\section{Introduction}
The rich spectrum of Lyman-alpha absorption lines in quasars (generally
known as the Lyman forest) was first noted by Lynds (1971) and studied in
detail by Sargent \etal\ (1980).  Because of its ubiquity and spatial
correlation properties it is usual to assume that the forest is a
distributed intergalactic population rather than being associated with
large galaxies.  In the first paper of the present series (Cowie \etal\
1995) we analyzed the metallicity of the Lyman forest in clouds with
column densities $\nhi\sim3\times\ten{14}\cmsq$ and found that these
objects have average metallicities of approximately $\ten{-2}$ solar and
consist of blends of much narrower components.  In the present paper we
turn to an analysis of the structure of lower column density forest clouds
which, of necessity, must be based on studies of the Lyman absorption lines
themselves since we do not currently have the sensitivity to detect metal
lines in these objects.

Profile fitting to high resolution spectra has been used to study the
velocity width (Doppler parameter $b$) and the neutral hydrogen column
densities $N$ of the clouds.  The distribution of Doppler parameters for
the \la\ forest was found to peak near $30\kms$ with a rather large spread
from 10 to $45\kms$ (Carswell \etal\ 1984; Atwood \etal\ 1985; Carswell
\etal\ 1987). Subsequently, on the basis of the very high resolution echelle
observations which had become possible by the end of the 1980's Pettini
\etal\ (1990) claimed a tight correlation between $b$ and $N$ with many
clouds having narrow profiles, leading to a median Doppler parameter of only
$17\kms$.  This result created understandable excitement in theoretical
circles since it seemed to question the standard picture of the
intergalactic matter confined in warm, extended, highly ionized gas clouds
rather than small, cool, almost neutral entities.  However, work by Carswell
\etal\ (1991) using an identical observational setup did not confirm the
Pettini \etal\ conclusions.  In a recent detailed analysis Rauch \etal\
(1993) showed that the profile fitting methodology is sensitive to noise
and that this can produce the results seen by Pettini \etal

The advent of the 10 m telescopes has resulted in an enormous improvement
in the S/N which can be achieved in high resolution spectra of the faint
quasars and in the present paper we discuss the distribution of low
column density forest clouds and the $b-N$ relationship based on such
data obtained with the HIRES spectrograph on the Keck telescope.  We
find that the $b$ distribution is quite invariant as a function of column
density and that there is a sharp cutoff in the number of clouds below
a value of about $b_c=20\kms$, corresponding to temperature $T=24,000$ K
for thermal broadening.  We then argue that the broader absorption lines
are blends of components, each with thermal broadening $b_c$, and each
broad line having an average of 3.25 components and a velocity dispersion
of $10.75\kms$.  This structure agrees with that found from the metal
lines in the absorption systems just above the present column density range.

\section{Data}

As part of a program to measure (D/H) ratios in high-$z$ absorption systems,
a number of bright ($V < 18$) $z\sim3$ quasars with partial Lyman-limit edges
(optical depth $\tau<3.5$) have been observed with the HIRES spectrograph
(Vogt \etal\ 1994) on the Keck 10 m telescope. A more complete description
of the data and data reduction procedures is given in Songaila (1995).
Each observation consists of a $3-6$ hr exposure covering the wavelength
range $3600\ang\to 6000\ang$ at a resolution $R$ of 36,000.  The wavelength
coverage is complete at shorter wavelengths ($\lambda\approxlt5000$\ang),
but has small interorder gaps at longer wavelengths.  Four of these spectra
(toward the quasars 0014+813, 0302--003, 0636+680, and 0956+122) are well
suited to a study of the low column-density ($\nhi\sim\ten{12}-\ten{14}$
cm$^{-2}$) \la\ forest lines since they have good S/N ($\approxgt50$ per
resolution element) and fully cover the wavelength range between the quasar
\la\ and \lb\ lines.  The properties of the quasars are summarized in Table 1.

For each of these four quasars a 600 \AA\ region of wavelength space was
chosen which lies above the wavelength at which \lb\ would enter the
spectrum but well below the quasar \la\ emission wavelength to avoid
proximity effects.  The wavelength intervals are summarized in Table 1.
The blaze function of the spectrograph in each order was removed by
normalizing to fits to white dwarf standard stars obtained in the same
configuration before and after each quasar observation.  The continuum
level of the spectrum was then established by an iterative fit to each
spectral order: all points above a critical value were fitted by a
third-order polynomial and renormalized; the critical value was then
raised until the extrema in the noise in the Lyman forest region were
matched to the extremal noise characteristics in the region above the
rest-frame Lyman alpha emission taking into account the system response
determined from the white dwarf observations and the overall spectral
shape of the quasar.  The normalized spectra were then spliced together to
form a total spectrum.  Each spectrum was fitted with an automatic
Voigt-profile fitting procedure which added lines at each minimum in the
spectrum where the average absorption depth per resolution element,
$\tau$, exceeded 0.05, until a satisfactory fit was achieved, in which the
residuals were consistent with the expected noise levels.  A sample region
of a spectrum with the fits and residuals is shown in Fig.~1.  There were
1056 lines found with $\nhi \geq 2\times \ten{12} \cmsq$.  The number of
detected lines for each quasar is given in Table 1 together with the
median $b$ values in two column density ranges.
The complete line catalogs and spectra are given in the Appendix.
For Q0014+813 we compared the detected clouds in detail with Rauch \etal's
(1992) profile fitting analysis of this quasar based on lower resolution
($23\kms$) and lower S/N data.  The agreement was generally good and
almost always within Rauch \etal's error estimates for both $b$ and $N$.
Almost all of the narrow lines in Rauch \etal's sample are found to be
caused by scatter, which is consistent with their prediction.

We have limited our primary analysis to the intermediate column density range
$2\times\ten{12}\cmsq<\nhi<3\times\ten{14}\cmsq$ where uncertainties due to
saturation are small.  Measured column densities at $\nhi>3\times\ten{14}\cmsq$
are highly uncertain, but there are 66 clouds in the wavelength range with
$\nhi\geq 3\times\ten{14}\cmsq$.

A plot of the $b$ values of the fitted lines vs.\ column densities is
shown in Fig.~2 for $\nhi$ in the range $3\times\ten{12}\cmsq$ to
$3\times\ten{14}\cmsq$.  The lower limit is chosen to avoid the optical
depth selection effect against shallow lines, which is illustrated in Fig.~2
by the solid line, which corresponds to $\tau\sim0.05$, and to minimize
noise artifacts in the profile fitting.  We have separated Q0636+680
(Fig.~2$a$), which has a relatively large number of narrow lines, from
Q0014+813, Q0302--003, and Q0956+122, which have relatively few narrow lines.
In Q0636+680 17 of the 266 lines in the $3\times\ten{12}\cmsq\leq\nhi\leq
3\times\ten{14}\cmsq$ range have $b<14\kms$.  A detailed analysis of these
narrow lines (Table~2) shows that 13 of these can straightforwardly be
identified as metal lines based on their correspondance to a strong \la\ line
and to other narrow lines in the spectrum.  The remaining narrow lines are
quite likely to be unidentified metal lines.  However, the most important
point to draw from Q0636+680 is that narrow lines can easily be retrieved
from the spectrum if they are present, and that the absence of large numbers
of narrow lines in the three other quasars is not a selection effect.  The
ease with which narrow lines can be picked out is also illustrated in Fig.~1.

For Q0014+813, Q0302--003, and Q0956+122 (Fig.~2$b$) there are only 11
out of 670 clouds with $3\times\ten{12}\cmsq\leq\nhi\leq3\times\ten{14}\cmsq$
which have $b<14\kms$.  Eight of these are straightforwardly identified as
metal lines (Table 2).  Thus less than 1\% of the \la\ forest clouds have
$b<14\kms$.

The $b-N$ distribution of Fig.~2$b$ has a fairly well defined lower
envelope which we have outlined with the dotted line $b=16\kms +
4(\nhi/\ten{14}\cmsq)\kms$.  As we shall show in Fig.~3 the slope in this
lower extremum is primarily an artifact of the much larger number of
clouds at lower \nhi\ and the increased scatter in the $b$ determinations,
and the distribution is consistent with a sharp cutoff below $b=20\kms$
which is invariant with column density.

Turning to the distribution of column densities we show the raw distribution
of counts per unit column density and $z$, $\left({\displaystyle {dn}}\over
{\displaystyle{dN\,dz}}\right)$, for the four quasars in Fig.~3.  The counts
are indistinguishable within the Poisson uncertainties, indicating that the
properties of the forest are highly invariant and that the retrieval of the
lines is reproducible from system to system.

\section{Analysis}

In analyzing the data of \S2 it is necessary to deal with the problems of
blending and selection since confusion becomes significant at the low
column density end.  Because there are a very large number of weak
clouds the probability of cloud blending is high and can result in
substantial incompleteness in the low column density counts and widening
of the $b$ value distribution.  Very faint clouds may also be lost if
their $b$ values are too large and their central optical depths are too
low compared to the noise levels.  This latter selection effect, which
is relatively unimportant at $\nhi>3\times\ten{12}\cmsq$, is illustrated
by the solid line in Fig.~2.

In order to remove these effects we have performed a number of simulations
to measure the cloud incompleteness and $b$ distribution widening as a function
of column density.  In these simulations we assumed that the distribution
of column densities was described by a $-1.5$ power law normalized to produce
the observed number of lines, and that the distribution of $b$ values was
described by a Gaussian with $\bar{b}=28\kms$ and $\sigma(b)=10\kms$ truncated
below $b=20\kms$.  We also assumed that the $z$ distributions were
uncorrelated.
These input parameters were chosen to produce output spectra which closely
match the observed spectra.  Artificial spectra were generated and convolved
with the instrumental spectral response with the appropriate Gaussian noise
added.  The finding program was then run on these artificial spectra to
obtain lists of clouds.

The completeness of the cloud recovery in the artificial spectra is
tabulated in Table 3.  Above $\ten{13} \cmsq$ essentially all clouds are
recovered, together with a small number of blends of lower column density
clouds,  but below this column density the cloud recovery rate drops, falling
to a completeness of only 25\% in the interval  $\nhi = 2\times \ten{12}
\cmsq\to4\times\ten{12}\cmsq$.  It is important to emphasize that this
effect is not determined by the S/N since the incompleteness is almost wholly
dominated by confusion: the recovery rate is essentially the same for an
artificial spectrum with infinite S/N.

The number of clouds per unit $z$ and unit \nhi\ in the observed spectra
is shown in Fig.~3 before and after correcting for the incompleteness.
The corrected distribution in the range $2\times\ten{12}-3\times\ten{14}\cmsq$
is well fit by a power law of the form $N^{-\beta}$ where $N$ is the
\hi\ column density.  The best fit to the power law index $\beta$ is
$-1.46$ with a 95\% confidence range of $(-1.37\to-1.51)$.

Remarkably, the power law exponent of the spectrum fitted over this limited
column density agrees well with the power law fit determined over much
larger \nh\ ranges (Tytler 1987, Sargent \etal\ 1989).  In Fig.~4 we have
compared the present data with Petitjean \etal's (1993) summary of recent
data at higher column densities and with a power law fit with $\beta=-1.46$.
For consistency with their notation Fig.~4 shows the quantity
\begin{equation}
  {f(N) = {\displaystyle{m}\over\displaystyle{\Delta N \, \sum \Delta X}}}
\end{equation}
where $m$ is the number of clouds in the intervals $dX$ and $dN$ with
$X \equiv 0.5 [\,(1+z)^2 - 1\,]$. $f(N)$ is roughly a factor of 3.8 less than
$\left({\displaystyle{dn}\over\displaystyle{dN\,dz}}\right)$.
$\sum\Delta X = 7.37$ for the present sample.  We have tabulated $f(N)$
in Table 3 for the present data.   The two data sets agree extremely well
in the overlap region.  The present data set may also be used to check
Petitjean \etal's finding that there is a significant deficit of
clouds in the range $\ten{15}-\ten{17}\cmsq$ compared to the power law
extrapolation from higher and lower column densities.  The best power
law fit to $f(N)$ in the regime $2\times\ten{12}-3\times\ten{14}\cmsq$,
given by
\begin{equation}
  f(N) = 4.9\times\ten{7}\,N^{-1.46}
\end{equation}
with all quantities expressed in cgs units,
would predict 193 clouds with $\nhi>3\times\ten{14}\cmsq$ compared to the
70 observed, implying that there is indeed a significant deficit between
$3\times\ten{14}\cmsq$ and $\ten{17}\cmsq$.  Setting $\beta=-1.51$
only reduces the predicted number of clouds to 127.  Thus there is indeed
a significant steepening of the slope above $\nhi=3\times\ten{14}\cmsq$.

The key issues for the $b-N$ distribution is whether the rise of the
lower envelope in $b$ with $N$ is real and whether there are any clouds
with very large $b$ or whether these are primarily random blends.  We have
compared the observed distribution of $b$ values in the high ($\nhi=
3\times\ten{13}-3\times\ten{14}\cmsq$) and low ($\nhi=3\times\ten{12}-3
\times\ten{13}\cmsq$) column density regimes in Fig.~5.  The models
give an acceptable fit to both distributions, showing that a single
cutoff at $b=20\kms$ can be adopted at all column densities.  A better
fit is given by adopting $b=18\kms$ in the regime $3\times\ten{12}-3\times
\ten{13}\cmsq$ and $b=22\kms$ in the $3\times\ten{13}-3\times
\ten{14}\cmsq$ range, so a weak increase in the cutoff $b$ with $N$ may
be slightly favored but is not required.  The drop in $b$ with decreasing
$N$ may occur because weaker lines are more susceptible to having their
Doppler parameters underestimated, due to the finite signal-to-noise of
the data.

\section{Discussion}

Our first observational conclusion is that the number of \la\ forest
clouds rises as a smooth $\beta=-1.5$ power law in the range
$2\times\ten{12}-3\times\ten{14}\cmsq$ and shows no sign of turnover at
the low column density end.  Confusion makes it essentially impossible to
extend this result to much lower column densities.  The second result is
that the forest clouds have a well defined lower envelope to the $b$
values of $b=20\kms$.  It is possible for this limit to have a weak
dependence on column density rising from $b=18\kms$ at
$3\times\ten{12}-3\times\ten{13}\cmsq$ to $b=22\kms$ at
$3\times\ten{13}-3\times\ten{14}\cmsq$, but this is not required by the
data.  The distribution of $b$ values can be adequately described by a
Gaussian with $\bar{b}=28\kms$ and a $\sigma(b)=10\kms$ truncated below
$b=20\kms$ at all measured column densities.

The mean intergalactic density, $n$, contributed by the clouds with these
column densities may be obtained by integrating equation (2) whence
\begin{equation}
  n={\displaystyle\hnought\over\displaystyle c}\,(1+z)^3\ (1+2\,\qnought\,
    z)^{1\over2}
    \int^{N_{max}}_{N_{min}} {\displaystyle {f(N)\, N\, dN} \over \displaystyle
    F(N)}\ .
\end{equation}
Here $F(N)$ is the mean ionization fraction at column density $N$.  Assuming
that $F$ is not such a steep function of $N$ that it weights the primary
density contribution to the low column density end, we can integrate
equation (3) to obtain
\begin{equation}
  \Omega_b = {{\displaystyle 1.5\times\ten{-7}}\over{\displaystyle <F>}}\,
  \left({\displaystyle N_{max} \over \displaystyle 3\times\ten{14}\cmsq}\right)
  ^{0.54}
\end{equation}
where $F$ is the appropriately weighted mean value and we have adopted
$\hnought=100\kms$ Mpc$^{-1}$ and \qnought=0.5.  The corresponding mean
physical separation is
\begin{equation}
  L=120~{\rm kpc}\ \left( {\displaystyle{N_{min}}\over
\displaystyle{2\times\ten{12}\cmsq}}\right) ^{0.46}.
\end{equation}

As Press \& Rybicki (1993) have emphasized, equation (4) shows at once
that the larger $b$ values cannot be caused by the thermal broadening
in the most simple photoionization model where the clouds are in thermal
equilibrium and have no kinematic broadening.  Thus from Donahue \& Shull
(1991)
\newcounter{saveeqn}
\newcommand{\alpheqn}{\setcounter{saveeqn}{\value{equation}}%
\stepcounter{saveeqn}\setcounter{equation}{0}%
\renewcommand{\theequation}{\mbox{\arabic{saveeqn}\alph{equation}}}}
\newcommand{\reseteqn}{\setcounter{equation}{\value{saveeqn}}%
\renewcommand{\theequation}{\arabic{equation}}}
\alpheqn
\begin{eqnarray}
  b\, & = & 28\kms\,U^{0.076}\\
  T & = & 4.9\times\ten{4}\,U^{0.152}\\
  F & = & 3.4\times\ten{-6}\,U^{-1.066}
\end{eqnarray}
\reseteqn
in terms of their ionization parameter $U$.  However, from equation (4)
we see that $<F>$ must be greater than $1.1\times\ten{-5}$ in order not
to exceed the baryon limit from nucleosynthesis of $0.01$
(Walker \etal\ 1991) and hence $U<0.3$ implying $b<25\kms$.

The structure of the \civ\ lines at slightly higher column densities
(Cowie \etal\ 1995) suggests an alternative picture. If we assume instead
that each cloud is a blend of comparable column density components and
that the observed $b$ value is produced by this blend of components each
with a thermally broadened $b_c$ and a centroid velocity spread described
by $f(v_c)$, then the value $b_c$ must be identified with the lower cutoff
in $b$ which will occur when an absorption line is dominated by a single
component.  This model is simplistic in assuming that all components have
similar column densities and that there is a single thermal broadening for
the clouds when in fact there are likely to be density and ionization
parameter variations from component to component.  However, the use of a
single value for $b$ is justified by the extremely weak dependence of $b$
on $U$ in equation (6).  Conversely, the extremely weak dependence of
$b_c$ on $U$ also implies that $U$ and $F$ can be determined only to order
of magnitude.  Thus for $b_c =20\kms$, $U=\ten{-2}$ and
$F=3.8\times\ten{-4}$, while for $b_c=18\kms$, $U=\ten{-2.5}$ and
$F=1.7\times\ten{-3}$, and for $b_c=22\kms$, $U=\ten{-1.4}$ and
$F=\ten{-4}$.   However, and very reassuringly, this agrees with the $U$
of $\ten{-2}\to\ten{-1.5}$ estimated by Cowie \etal\ (1995) based on the
ionization balance in clouds just above $\nhi=3\times\ten{14}\cmsq$.
Furthermore, the hydrogen thermal broadening would predict a thermal $b$
value of $6\kms$ for carbon lines, which is very close to the value of
$7\kms$ found by Cowie \etal\ \ Finally, with $F=3\times\ten{-4}$, then
$\Omega_{b}=5\times\ten{-4}$ and very roughly 4\% of the baryons inferred
from nucleosynthesis lie in these components.  Thus this model appears to
provide a self-consistent description.

We can determine the average number of components and their velocity
spread by comparing the predicted spread in $b$ values with the
observed distribution.  If we assume that each absorption line consists
of a number of components $n$, each thermally broadened with $b_c=20\kms$
and described by the probability distribution
\begin{equation}
  P(n) \sim{\displaystyle 1\over\displaystyle n!}\, ({\rm N})^n\quad
  {\rm with~} n\geq1
\end{equation}
and that the centroids of these velocity components are described by a
Gaussian velocity distribution function characterized by $\sigma(v_c)$,
then we can model the observed $b$ value distribution function to determine
N and $v_c$.  We find that these quantities are extremely tightly
constrained with a best fit of N=3.1, corresponding to a mean of 3.25
components per absorption line, and $\sigma(v_c)=10.75\kms$.  Any smaller
value of N results in too many single component clouds with $b=b_c$ while
larger N result in a Gaussian spread of $b$ values which is also a poor
fit to the observed distribution.  The velocity spread is slightly smaller
than the value of $18\kms$ found by Cowie \etal\ (1995) for the spread in
\civ\ components in $\nhi\geq3\times\ten{14}\cmsq$ clouds and the mean
number of \civ\ components (2.5) per absorption line systems is slightly
smaller.  This could represent a velocity dependent ionization effect or a
difference between the column density samples, but most likely it is due
to observational uncertainty and reflects the difficulty of measuring the
\civ\ dispersion.  The presently determined dispersion should be
considerably more accurate.

At larger velocity separations the components are no longer blended and
we may directly investigate the two-point correlation function in the form
\begin{equation}
  \xi(v)={\displaystyle N_{pairs}(v) \over \displaystyle N_{pairs}^{art}(v)}
  -1
\end{equation}
where $N_{pairs}(v)$ is the number of observed pairs at separation $v$
and $N_{pairs}^{art}(v)$ is the number of expected pairs determined from
the artificial spectra with no correlation.  This is illustrated in
Fig.~6.  For the forest sample with
$7\times\ten{12}\cmsq<\nhi<3\times\ten{14}\cmsq$ we find no evidence
of correlation at $v>150\kms$ consistent with previous studies (Sargent
\etal\ 1980) but for $50\kms<v<150\kms$ we do find a significant excess
of $0.17\pm0.045$ -- slightly smaller than, but consistent with, the
value of $0.32\pm0.08$ found by Webb (1987).  Intriguingly, if we divide
the sample by column density the correlation increases with a value of
$0.73\pm0.13$ for $4\times\ten{13}<\nhi<3\times\ten{14}\cmsq$ and
$0.21\pm0.07$ for $7\times\ten{12}<\nhi<4\times\ten{13}\cmsq$ suggesting
that clouds are preferentially correlated with clouds of similar density,
and that there might be a weakening of the correlation as we move to lower
column densities.  Crotts (1989), Chernomordik (1995), and Cristiani \etal\
(1995) have previously noted such an effect.  Cristiani \etal\ found a
correlation with a value of $0.89\pm0.18$ at $\Delta v\leq100\kms$ for
$\nhi\geq6\times\ten{13}\cmsq$ clouds at slightly higher redshift in the
spectrum of the quasar Q0055--269.  For $2\times\ten{13}\cmsq\leq\nhi\leq
6\times\ten{13}\cmsq$ they obtain $0.38\pm0.1$.  These values are roughly
consistent with the present results, and together they may suggest that the
correlation function does weaken at lower column densities.  Suggestions of
anti-correlation features in the two-point correlation at a few to several
hundred km~s$^{-1}$ may also be noted in Fig.~6. While we cannot be sure of
the nature of this phenomenon, such features appear real.  Inspection of
figures 8 and 9 of Cristiani \etal\ (1995) shows that similar negative features
are visible in the correlation function of their high column density sample,
log~$\nhi>14$ (see also Meiksin \& Bouchet 1995). Anticorrelation features
are expected to occur when some \la\ clouds are forming velocity
caustics during turnaround (Kaiser 1987; McGill 1990), but on smaller
scales of a few hundred km~s$^{-1}$ they may be more likely to be due to
gas dynamical effects like pressure fluctuations (Webb \& Barcons
1991) or inhomogeneous ionization (Meiksin \& Bouchet 1995).

Finally, we may combine the present data with the properties inferred from the
metal lines in the $\nhi=3\times\ten{14}\cmsq$ \la\ forest to estimate the
cloud properties.  Cowie \etal\ (1995) found from the ionization balance
in the metals that $U=\ten{-2}\to\ten{-1.5}$ for $\nhi=3\times\ten{14}\cmsq$
clouds at this redshift, and, that the neutral hydrogen fraction was then
$4.6\times\ten{-4}\ U_{-2}{}^{-1.066}$ (from Eq. 6(c) where $U_{-2}\equiv
U/\ten{-2}$).  By the definition of $U$ the total density in the cloud is
$n_{tot}=6.3\times\ten{-3}\,J_{-21}/U_{-2}$ where $J_{-21}$ is the incident
ionizing intensity just above the Lyman edge in units of $\ten{-21}$ ergs
cm$^{-2}$ s$^{-1}$ Hz$^{-1}$ sr$^{-1}$.  Assuming this ionization fraction
holds over the relevant column density range, the line-of-sight cloud size,
$D = 10\ {\rm pc}\, U_{-2}{}^{2.066}\,J_{-21}{}^{-1}$ $(\nhi/\ten{14}\cmsq)$.
This is $\ten{-5}$ to $\ten{-4}$ of the separation distance for $\nhi=
\ten{14}\cmsq$, suggesting that the filling factor of the forest clouds is
correspondingly small.  The physical dimensions are similarly small compared
to the transverse dimensions (Bechtold \etal\ 1994; Dinshaw \etal\ 1994),
suggesting that the clouds are highly flattened.  (Rauch \& Haehnelt 1994
give a more indirect argument to this effect.)\ \ The internal density
represents an overdensity of $1400\left({\displaystyle\Omega_{IGM}\over
\displaystyle0.01}\right)^{-1}$ with respect to the average IGM, where
$\Omega_{IGM}$ is the fraction of the closure density in baryons.

The internal kinematic structure strongly favors models in which the
underlying formation mechanism is gravitational or kinematic such as the
minihalo model of Rees (1986) or the pancake model of Cen \etal\ (1994).
The objects are extremely elongated, which would seem to favor the latter
model.  These models allow a natural understanding of the kinematic
structure and the spread of velocities, since higher infall or shock
velocities would result in too high a temperature for neutral hydrogen to
be present.  Indeed, the pancake model produces a natural explanation for
the number of components, since we will see the pancake itself, the
outgoing shocks, and possibly the surrounding infalling material.  In its
current version the pancake model does not correctly predict the
extrapolation of the power law distribution of column densities below
$\ten{13}\cmsq$ found in the present paper but this may be very much a
function of the assumed details.

\appendix\vskip6ex plus .7ex
\centerline{\bf APPENDIX}

Plots and line lists for spectral line fits for Q0014+813, Q0302--003,
Q0636+680, and Q0956+122.  Each plot shows the normalized spectrum
with overlying profile fits, plotted against vacuum heliocentric
wavelength.  Vertical tick marks indicate the positions of line
features, with every tenth line numbered.  Lines with $\nhi\geq
3\times\ten{12}\cmsq$ and $b<14\kms$ are marked by the filled boxes.
Residuals from the fits are plotted below each spectrum.  A tabulation
of the line identifications by wavelength, with $b$ value, and redshift
and column density (under the assumption that these are \la\ features)
is given in Tables~A1 and A2.

\acknowledgments
We would like to thank Jerry Ostriker, Jordi Miralda-Escud\'e, and Martin
Haehnelt for very helpful discussions.  These observations would not have
been possible without Steve Vogt's HIRES spectrograph.

\newpage
\begin{figure}
\caption{A sample 50\ang\ region of the spectrum of Q0302--003 showing fits
to identified lines overlaid upon the normalized spectrum.  Residuals to
the fits are shown below, and it can be seen that there are no systematic
residuals associated with the profile fits.  The wavelengths of each line
are marked by ticks above the spectrum.  There are two lines with
$b<14\kms$ in this region of spectrum where the ticks are overlaid by
boxes.  These lines correspond to a \civ\ doublet at $z=1.8924$ (Table 2)
and are easily distinguished from the broader forest lines. The number of
lines fitted for each quasar is summarized in Table 1.}
\end{figure}
\begin{figure}
\caption{The distribution of components with respect to column density and
$b$ value for the quasar systems.  Q0014+813, Q0302--003, and Q0956+122
have been plotted separately from Q0636+680, since the latter system has
many narrow features known to correspond to metal lines.  For each figure
we have plotted only clouds with $3\times\ten{12}\cmsq< \nhi<3\times
\ten{14}\cmsq$ shown by the dashed lines.  Very broad clouds with low
column densities and high $b$ values will not be detected.  The solid line
shows this selection for $\tau=0.05$, above which clouds should be easily
found.  We have attempted to identify all cloud with $b\leq14\kms$ as
metal lines.  Such narrow components that have been identified with metal
line systems are indicated with a surrounding box.  The dotted line shows,
for Fig.~2b, the apparent lower bound to the $b$ values of \la\ forest
clouds.  The boundary appears to be a real physical limit, and the plot of
Q0636+680 shows that the fitting program is quite capable of picking out
narrow components where these are present.}
\end{figure}
\begin{figure}
\caption{The raw and corrected distributions of counts per unit
column density and redshift in the column
density regime $2\times\ten{12}-3\times\ten{14}\cmsq$.  Squares show the
distribution for Q0636+680, diamonds for Q0302--003, asterisks for
Q0014+813, and triangles for Q0956+122.  The dashed line shows the average
raw counts for all four quasars.  It can be seen that the raw counts are
indistinguishable from system to system, within the Poisson noise, and that
both the properties of the \la\ forest and the capacity of the program to
retrieve lines is highly invariant for all three quasars.  The large open
boxes show the column density distribution corrected for incompleteness
with the best power law fit shown as a solid line.}
\end{figure}
\begin{figure}
\caption{Comparison of the cloud number distribution at low and high
column densities.  The data in the present survey, covering the column
density range $2\times\ten{12}-3\times\ten{14}\cmsq$, are shown with
filled squares, along with the best power-law fit ($-1.46$) over this
region.  This fit has been extrapolated to higher column densities, and
the higher column density data compiled by Petitjean \etal\ (1993) is
shown with diamonds overlaid with the associated error bars.  The
deviations below the line around $\ten{15}-\ten{17}\cmsq$ which were noted
by Petitjean \etal\ are confirmed by the present data (see text).
However, as was originally emphasized by Tytler (1987), the overall fit to
a single power-law is remarkably good over nearly ten orders of magnitude
in column density.}
\end{figure}
\begin{figure}
\caption{Comparison of the distribution of $b$ values derived from the
artificial spectra compared with the observed distributions.  The lower
left panel shows the input distribution for the artificial spectra
averaged over eight realizations.  In the right hand panels we show the
recovered distributions (dashed lines) compared with the observed
distributions (solid lines) in the high and low column density ranges.
The observed distribution is the average for Q0014+813, Q0302--003,
and Q0956+122 with known metal lines excluded.}
\end{figure}
\begin{figure}
\caption{The correlation functions vs.\ velocity separation in the column
density ranges $4\times\ten{13}-3\times\ten{14}\cmsq$ (upper graph) and
$7\times\ten{12}-4\times\ten{13}\cmsq$ (lower graph).  The correlation
functions have been computed in $50\kms$ bins and the dotted line shows
the $1\sigma$ error computed from the number of artificial pairs.}
\end{figure}
\newpage
%
%
\begin{deluxetable}{lcrrcccc}
\tablewidth{42pc}
\tablecaption{Fitted Quasars}
\tablehead{
\colhead{}                   & \colhead{}                   &
\colhead{}                   & \colhead{}                   &
\colhead{}                   & \colhead{}                   &
\colhead{\chhlap{Low $N$}}   & \colhead{\chhlap{High $N$}} \\[0.5ex]
\colhead{}                   & \colhead{$m$}                &
\colhead{}                   & \colhead{$\lambda\lambda$}   &
\colhead{\chlap{No.\ lines}} & \colhead{\chlap{No.\ lines}} &
\colhead{\chhlap{\small Median$\,b$}} & \colhead{\chhlap{\small Median$\,b$}}
\\[0.5ex]
\colhead{Quasar}             & \colhead{\chhhlap{\small$(1450\ang)$}} &
\colhead{$z_{em}$}           & \colhead{(\AA)}              &
\colhead{\chlap{\small $N\!>2\times\ten{12}\cmsq$}}         &
\colhead{\chlap{\small $N\!>3\times\ten{14}\cmsq$}}         &
\colhead{\chhlap{\small (km s$^{-1}$)}}                     &
\colhead{\chhlap{\small (km s$^{-1}$)}}}

\startdata
\phd Q$0014+813$ & 16.7 & 3.384 & 4500--5100 & {\hbox to0.9in{\hfil254\hfil}} &
{\hbox to 0.9in{\hfil17\hfil}}~~~ & {\hbox to 0.4in{\hfil30\hfil}} &
{\hbox to 0.4in{\hfil32\hfil}}\nl
\phd Q$0302-003$ & 17.8 & 3.286 & 4400--5000 & 258  & 12~~~ & 31 & 32\nl
\phd Q$0636+680$ & 16.5 & 3.174 & 4300--4900 & 302  & 19~~~ & 26 & 24\nl
\phd Q$0956+122$ & 17.8 & 3.301 & 4400--5000 & 242  & 18~~~ & 30 & ~29%
\tablecomments{ For each quasar, the number of lines fitted and median $b$
values are given for the indicated
wavelength intervals in the observed frame.  The number of lines
fitted by the program is given for systems with column densities in
excess of $\nhi=2\times\ten{12}$ cm$^{-2}$ and $\nhi=3\times\ten{14}\cmsq$.
Median $b$ values are given for the column density ranges
$\nhi=3\times\ten{12} -3\times\ten{13}\cmsq$ and
$\nhi=3\times\ten{13}-3\times\ten{14}\cmsq$. Quasar magnitudes and
emission-line
redshifts are taken from Sargent \etal\ (1989).}
\enddata
\end{deluxetable}

\clearpage
%
%
\begin{deluxetable}{lccl}
\tablewidth{26pc}
\tablecaption{Narrow Line Identifications}
\tablehead{
\colhead{}                   & \colhead{$\lambda$}    &
\colhead{$b$}                & \colhead{}\\[0.5ex]
\colhead{~~\ Quasar}         & \colhead{(\AA)}        &
\colhead{\chhlap{~~\small (km\thinspace{s}$^{-1}$)}}  & \colhead{ID}}
\startdata
{}~~Q$0014+813$ & 4538.65 & \phn9 & \cii\ (1334.5) 2.4009\nl
              & 4582.39 &    10 & \omit\hfil---\hfil\nl
              & 4597.20 & \phn7 & \omit\hfil---\hfil\nl
{}~~Q$0302-003$ & 4478.01 &    13 & \civ\ (1548.2) 1.8924\nl
              & 4485.45 &    11 & \civ\ (1550.8) 1.8924\nl
              & 4832.47 &    14 & \alii\ (1670.8) 1.8923\nl
{}~~Q$0636+680$ & 4380.44 & \phn7 & \silii\ (1260.4) 2.4754\nl
              & 4390.19 &    14 & \omit\hfil---\hfil\nl
              & 4417.85 & \phn2 & \cii\ (1334.5) 2.3104\nl
              & 4494.69 &    10 & \civ\ (1548.2) 1.9032\nl
              & 4537.95 & \phn2 & \civ\ (1548.2) 1.9311\nl
              & 4545.47 & \phn7 & \civ\ (1550.8) 1.9310\nl
              & 4545.81 & \phn7 & \civ\ (1550.8) 1.9313\nl
              & 4637.98 & \phn3 & \cii\ (1334.5) 2.4754\nl
              & 4642.63 &    10 & \omit\hfil---\hfil\nl
              & 4646.95 &    11 & \siliii\ (1190.2) 2.9043\nl
              & 4648.28 & \phn3 & \omit\hfil---\hfil\nl
              & 4694.16 &    10 & \omit\hfil---\hfil\nl
              & 4811.25 & \phn2 & \mgii\ (2795.5) 0.7211\nl
              & 4843.76 & \phn9 & \civ\ (1548.2) 2.1287\nl
              & 4848.16 &    10 & \civ\ (1548.2) 2.1315\nl
              & 4851.89 & \phn8 & \civ\ (1550.8) 2.1287\nl
              & 4856.22 &    11 & \civ\ (1550.8) 2.1315\nl
{}~~Q$0956+122$ & 4613.92 & \phn6 & \siliv\ (1393.8) 2.3104\nl
              & 4614.65 & \phn8 & \siliv\ (1393.8) 2.3109\nl
              & 4644.50 & \phn8 & \siliv\ (1402.8) 2.3109\nl
              & 4783.12 &    11 & \omit\hfil---\hfil\nl
              & 4963.75 & \phn8 & \siliii\ (1206.5) 3.1141\nl
\enddata
\end{deluxetable}

\clearpage
%
%
\begin{deluxetable}{lccc}
\tablewidth{26pc}
\tablecaption{Column Density Distribution}
\tablehead{
\colhead{Log$_{10}$ N} & \colhead{} &
\colhead{}             & \colhead{}\\[0.5ex]
\colhead{Range}        & \colhead{Incompleteness} &
\colhead{No.\ clouds}  & \colhead{Log$_{10}$ $f$(N)}}
\startdata
12.30--12.60 & 0.25 & 109 & --10.53\nl
12.60--12.90 & 0.52 & 188 & --10.91\nl
12.90--13.20 & 0.82 & 197 & --11.39\nl
13.20--13.51 & 1.03 & 173 & --11.85\nl
13.51--13.81 & 1.11 & 133 & --12.29\nl
13.81--14.11 & 1.08 & \phn88 & --12.76\nl
14.11--14.41 & 1.01 & \phn75 & --13.10\nl
$>14.41$     & 1.0  & \phn77 & ---\nl
\enddata
\end{deluxetable}

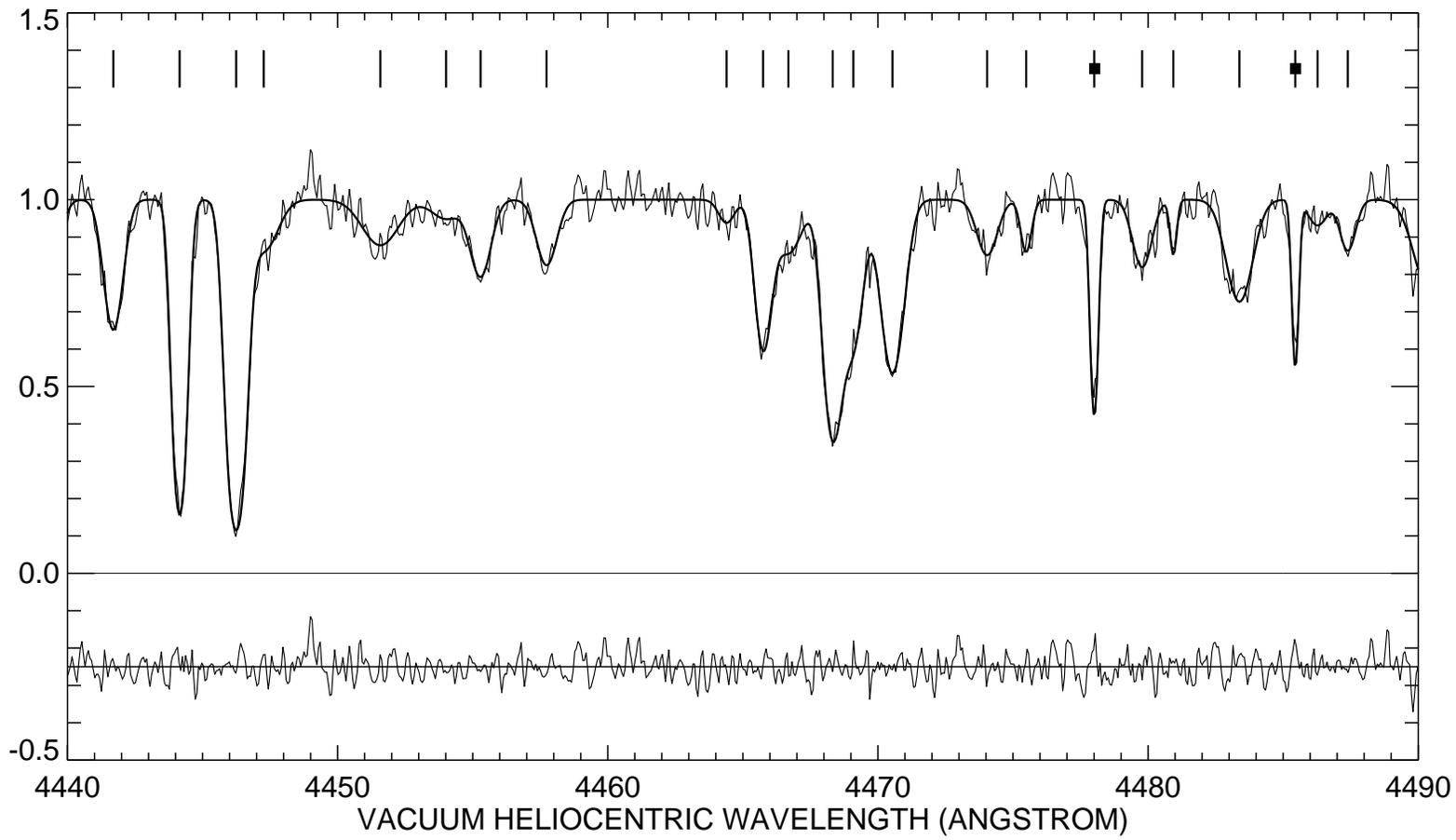
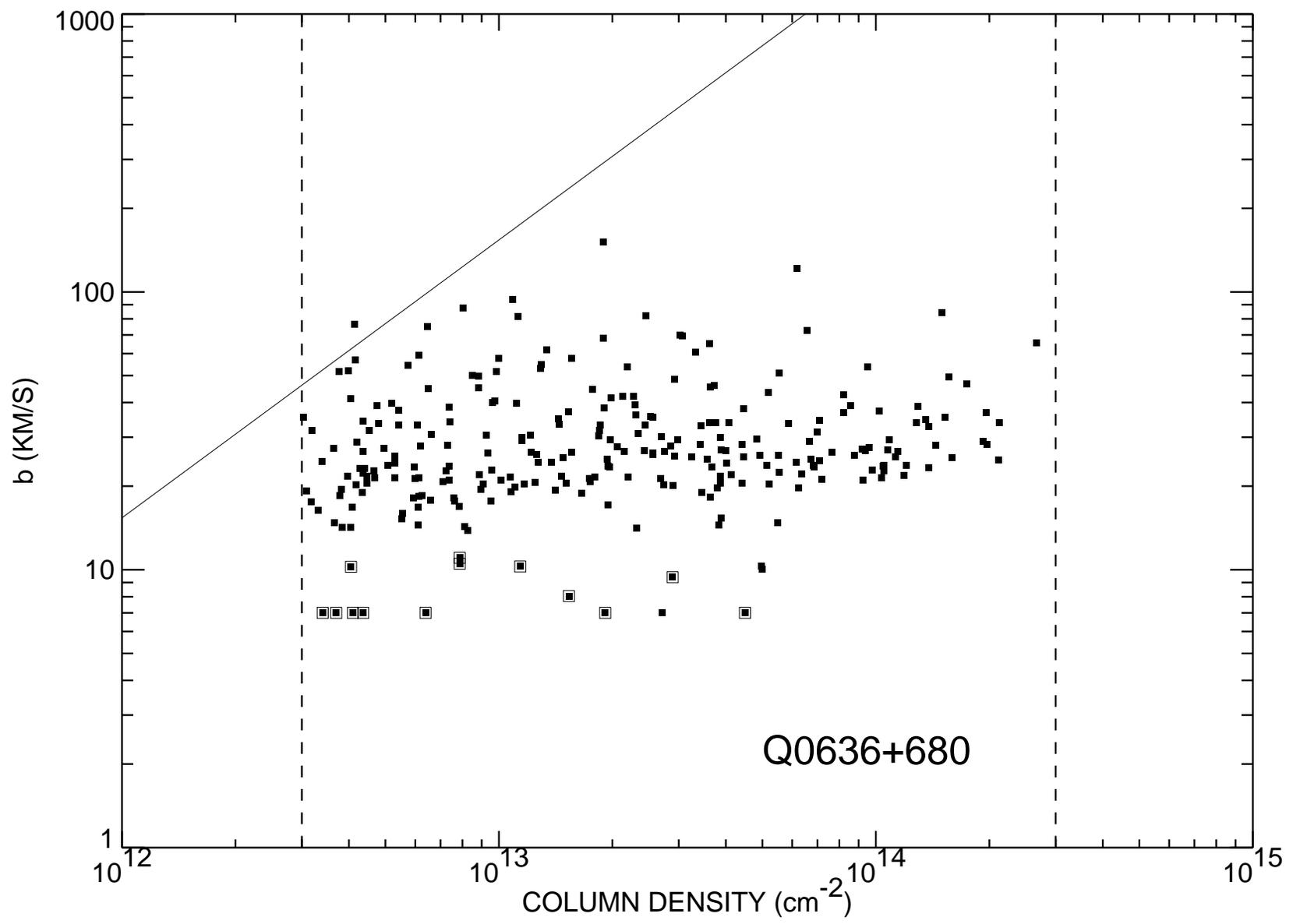
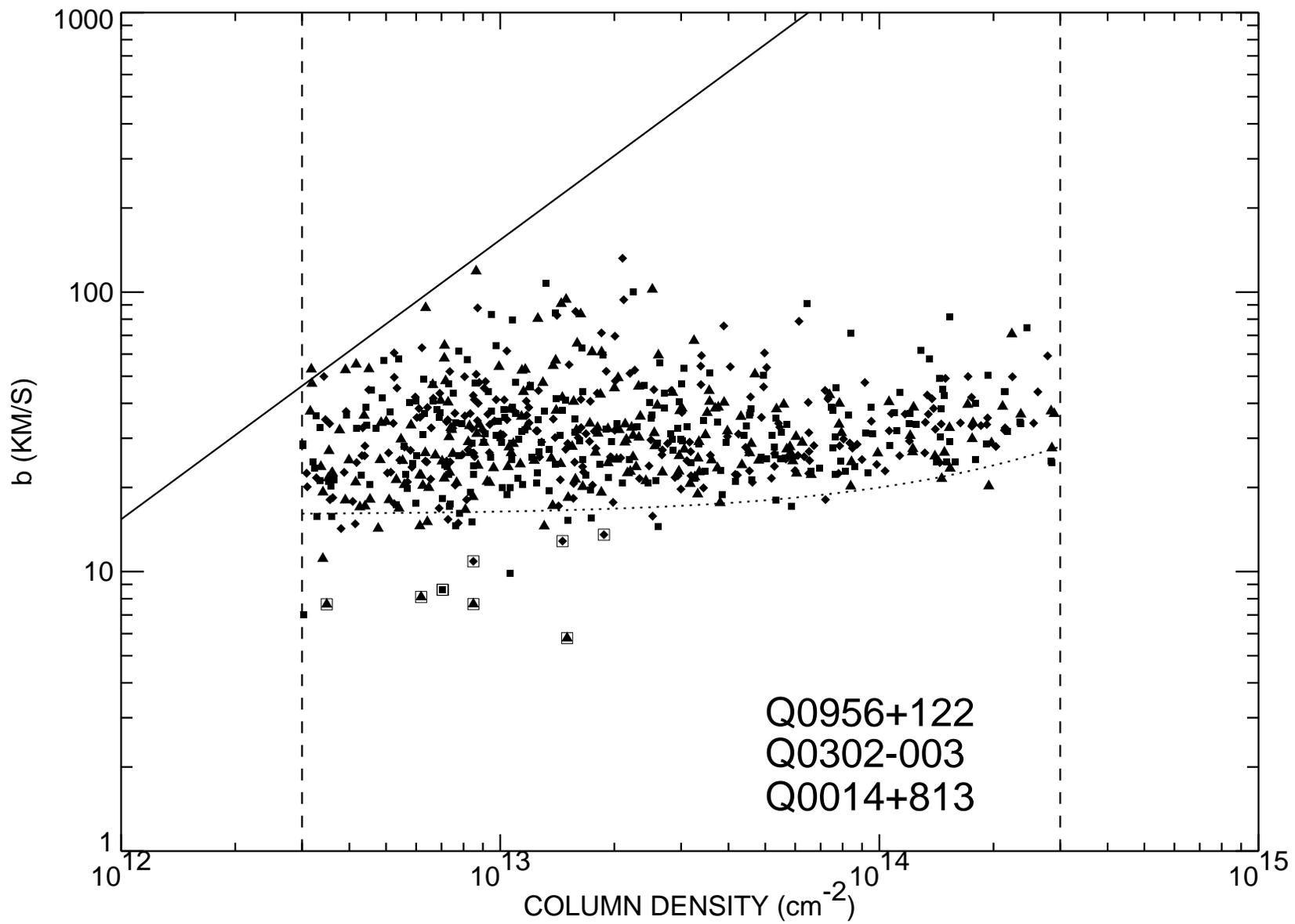
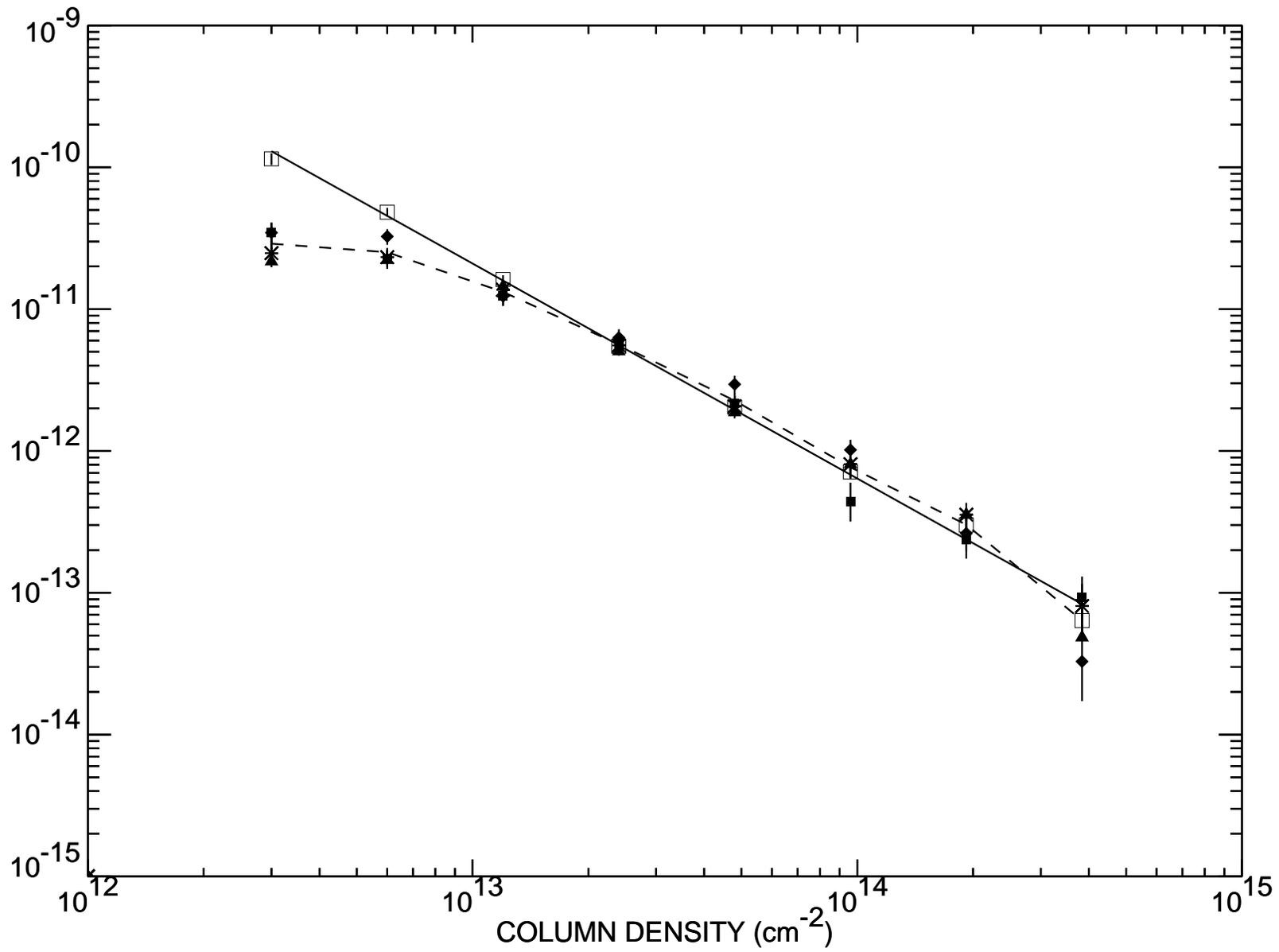
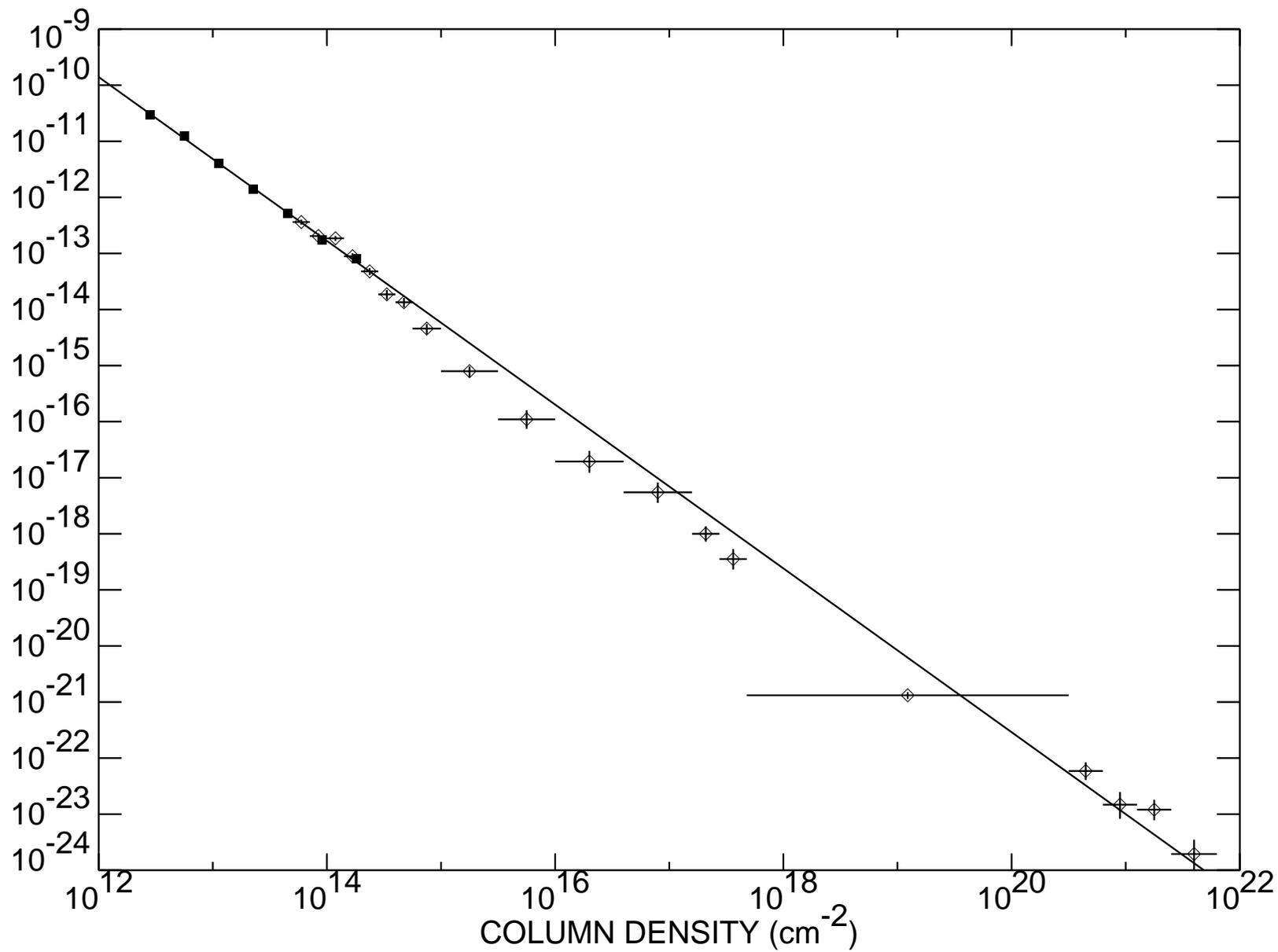
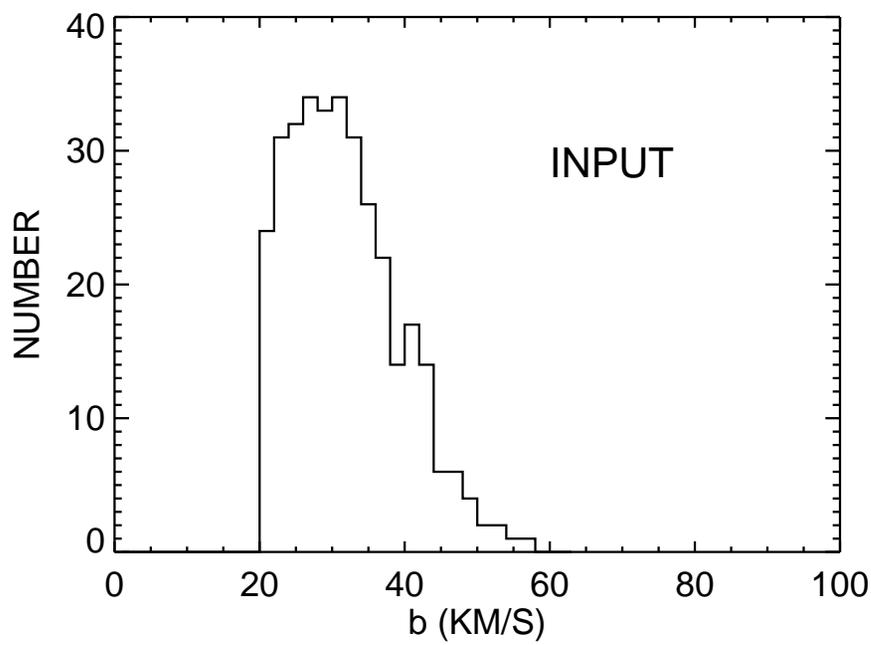
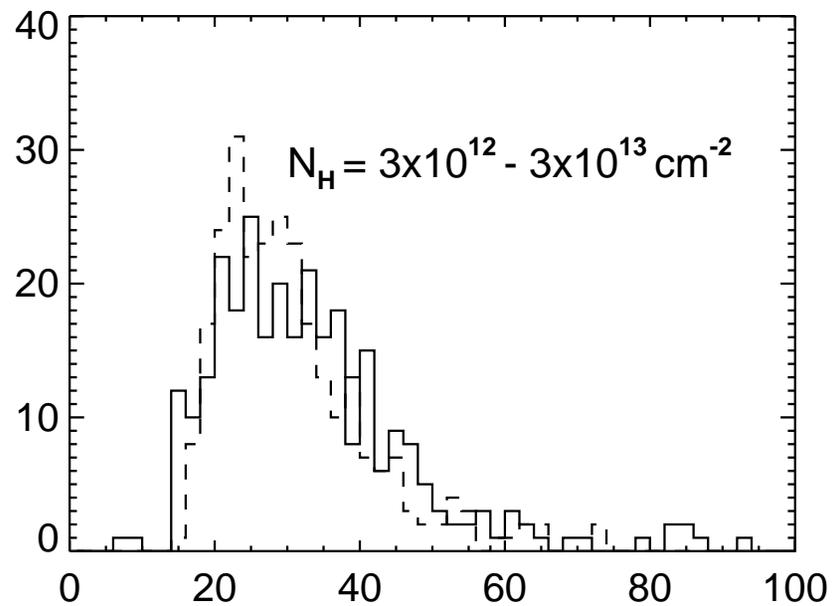
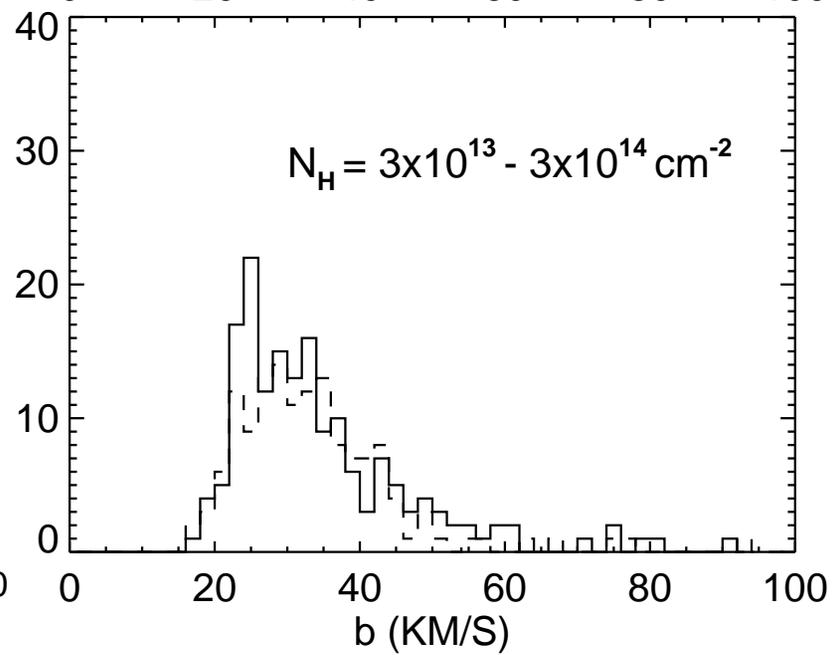
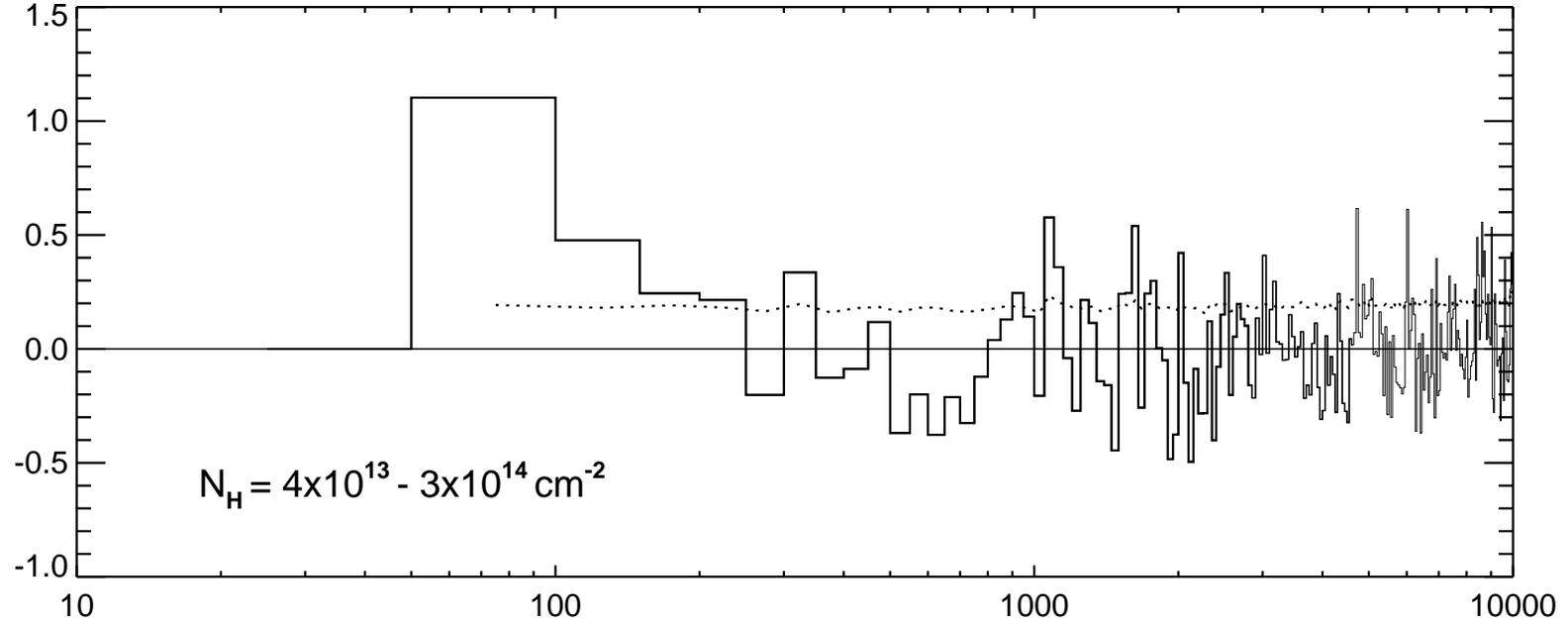
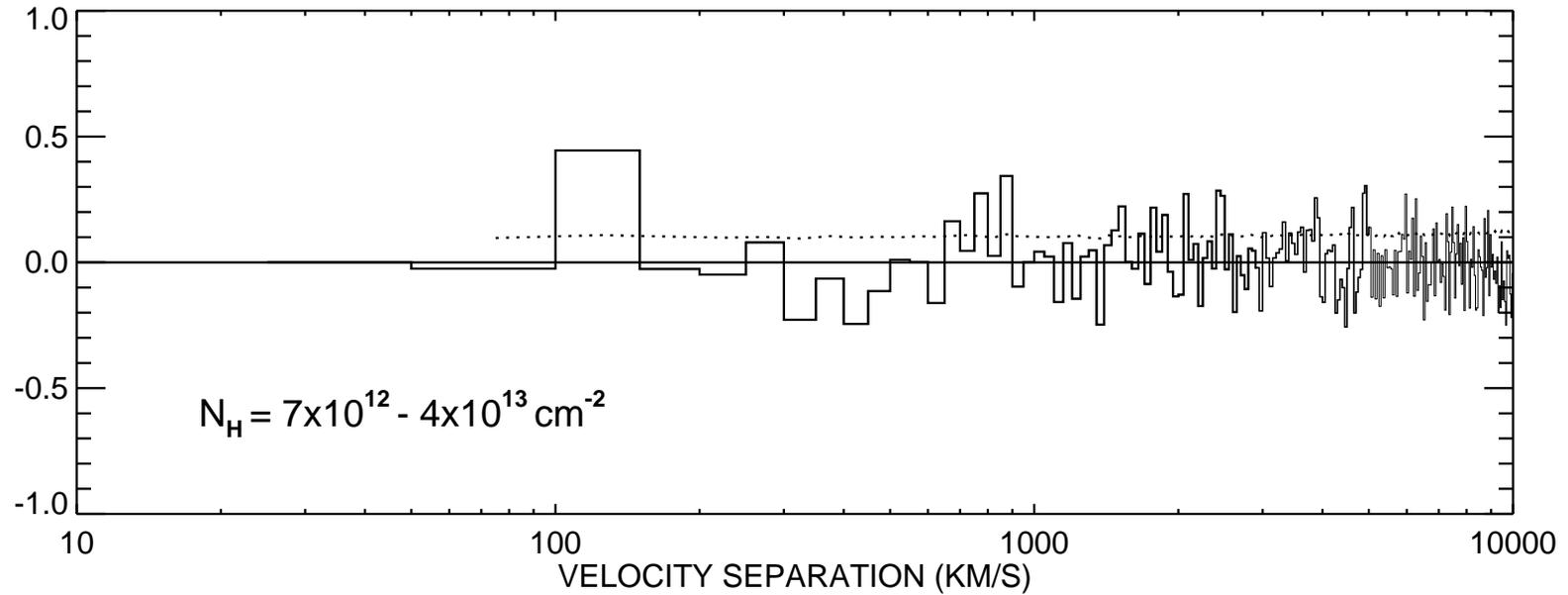
\end{document}